\documentclass[preprint]{aastex}
\slugcomment{To appear in the 10 March 2006 issue (vol.\ 639) of the {\it ApJ}}
\shortauthors{Kirkpatrick}
\shorttitle{Young Brown Dwarf}

\begin{document}

\title{Discovery of a Very Young Field L Dwarf, 2MASS J01415823$-$4633574\footnote{Some of the data presented herein were obtained at the W.~M.\ Keck Observatory, which is operated as a scientific partnership among the California Institute of Technology, the University of California, and the National Aeronautics and Space Administration. The Observatory was made possible by the generous financial support of the W.~M.\ Keck Foundation.}}

\author{J.\ Davy Kirkpatrick}
\affil{Infrared Processing and Analysis Center, MS 100-22, California 
    Institute of Technology, Pasadena, CA 91125; davy@ipac.caltech.edu}

\author{Travis S.\ Barman}
\affil{Department of Physics and Astronomy, University of California, Los Angeles, CA 90095-1562}

\author{Adam J.\ Burgasser\footnote{Visiting Astronomer at the Infrared  
    Telescope Facility, which is operated by the University of Hawaii under 
    Cooperative Agreement no.\ NCC 5-538 with the National Aeronautics and 
    Space Administration, Office of Space Science, Planetary Astronomy 
    Program.}}
\affil{Massachusetts Institute of Technology, 77 Massachusetts Avenue, Building 37, Cambridge, MA 02139}

\author{Mark R.\ McGovern}
\affil{Antelope Valley College, 3041 West Avenue K, Lancaster, CA 93536}

\author{Ian S.\ McLean}
\affil{Department of Physics and Astronomy, University of California, Los Angeles, CA 90095-1562}

\author{Christopher G.\ Tinney}
\affil{Anglo-Australian Observatory, P.O.\ Box 296, Epping, NSW 1710, Australia}

\author{Patrick J.\ Lowrance}
\affil{Spitzer Science Center, MS 220-6, California Institute of Technology, Pasadena, CA 91125}

\begin{abstract}

While following up L dwarf candidates selected photometrically from the
Two Micron All Sky Survey, we uncovered an unusual object designated 2MASS J01415823$-$4633574. 
Its optical spectrum exhibits very strong bands of vanadium 
oxide but abnormally weak absorptions by titanium oxide, potassium, and sodium. Morphologically such spectroscopic
characteristics fall intermediate between old, field early-L dwarfs 
($log(g){\approx}5$) and very late
M giants ($log(g){\approx}0$), leading us to favor low gravity as the explanation for the unique spectral signatures of this L dwarf. Such a
low gravity can be explained only if this L dwarf is much lower in mass than a typical old field L dwarf of similar temperature and is still contracting to its final radius. These conditions imply a
very young age. Further evidence of youth is found in the near-infrared spectrum, including a triangular-shaped H-band continuum reminiscent of young brown dwarf candidates 
discovered in the Orion Nebula Cluster. 
Using the above information along with comparisons to brown dwarf
atmospheric and interior models, our current best estimate is that this L dwarf 
has an age of 1-50 Myr and a mass of 6-25 $M_{Jupiter}$. 
Although the lack of a lithium detection (pseudo equivalent width
$<$1 \AA) might appear to contradict other evidence of youth, we suggest that lithium becomes weaker
at lower gravity like all other alkali lines and thus needs to be carefully considered 
before being used as a diagnostic of age or mass for objects in this regime. The location of 2MASS 0141$-$4633
on the sky coupled with a distance estimate of $\sim$35 pc and the above age estimate suggests that this object may be a brown dwarf
member of either the 30-Myr-old Tucana/Horologium Association or the $\sim$12-Myr-old $\beta$ Pic Moving Group. Distance as 
determined through trigonometric parallax (underway) and a measure of the 
total space motion are needed to test this hypothesis.

\end{abstract}

\keywords{brown dwarfs, young associations, L dwarfs, 
2MASS J01415823$-$4633574}

\section{Introduction}

The search for dwarfs later than late-M has had a brief, but surprisingly productive, history. The first L dwarf was discovered during a search for brown dwarf companions to nearby (white dwarf) stars by \cite{becklin1988}. Now the total number of L dwarfs recognized in the Solar Neighborhood exceeds 400 and the number of their cooler brethren, the T dwarfs, exceeds sixty\footnote{See the compilation at \url{http://www.DwarfArchives.org}.}. Research on L and T dwarfs is a rapidly maturing field, and having such a large number of cool objects enables studies that might not otherwise be possible \citep*{kirkpatrick2005araa}.

One such study is the identification of L and T dwarfs that show rare spectroscopic traits. A few notable L and T dwarfs with unusual or unique spectroscopic features have been identified. 
Explanations for the unusual features vary -- 2MASS J05185995$-$2828372 is believed to be a close binary composed of a late-L and mid-T dwarf \citep*{cruz2004} whereas 2MASS J05325346+8246465 is probably a low metallicity L dwarf member of the halo \citep*{burgasser2003}. In this paper we present the discovery of a field L dwarf whose unique spectroscopic signatures are likely caused by low gravity, which implies that it is a very young, low-mass brown dwarf.

\section{Discovery and Data Acquisition}

2MASS J01415823$-$4633574 (hereafter abbreviated as 2MASS 0141$-$4633) was discovered during a photometric search for L dwarfs using data from the 
Two Micron All Sky Survey (2MASS, \citealt{skrutskie2005}). The goal of this search was to produce a sizable list of southern hemisphere targets whose trigonometric parallaxes could be obtained
from the Anglo Australian Observatory. 2MASS 0141$-$4633 was selected because of its red
color ($J-K_s=1.735{\pm}0.054$) and relatively bright magnitudes of 
$J=14.832{\pm}0.043$, $H=13.875{\pm}0.026$, and $K_s=13.097{\pm}0.032$ as 
measured by the 2MASS All Sky Point Source Catalog \citep*{cutri2003}. Figure 1
shows optical and near-infrared finder charts of the region.
Additional information on other objects discovered during this L dwarf search 
can be found in 
\cite{kirkpatrick2005}.

\subsection{Keck LRIS Data}

2MASS 0141$-$4633 was followed up spectroscopically using the Low Resolution Imaging Spectrograph \citep*{oke1995} at the 10-m 
W.~M.~Keck Observatory on the clear night of 2003 Dec 24 UT. A 400 lines/mm grating
blazed at 8500 \AA\ was used with a 1$\arcsec$ slit and a 2048$\times$2048
CCD to produce spectra covering the range
6300 -- 10100 \AA\ and having a resolution of 9 \AA.  The OG570 order-blocking filter was used
to eliminate second-order light.  Flat-field exposures of the interior of the 
telescope dome were used to normalize the response of the detector. 
2MASS 0141$-$4633 was observed in two 1200-second exposures. Between these two exposures, a 20-second
spectrum of the G0 dwarf
HD 11342 -- 1.5 degrees away on the sky -- was observed to provide
a template for telluric correction. The data, all taken at the parallactic angle to mitigate differential color refraction, were reduced and calibrated 
using standard IRAF routines as detailed in \cite{kirkpatrick1999}. 
We interpolated the continuum across the 
O$_2$ bands at 6867-7000 and 7594-7685 \AA\ and the H$_2$O bands at 7186-7273,
8161-8282, $\sim$8950-9300, and $\sim$9300-9650 \AA\ in the resulting spectrum
of HD 11342
to produce a ``telluric free'' version. This cleaned spectrum of the G dwarf was then divided
into the pre-cleaned spectrum of itself to produce a telluric correction factor as a function of wavelength. The reduced spectrum of 
2MASS 0141$-$4633 was then multiplied by this correction factor to produce a telluric-corrected version of the target.

\subsection{IRTF SpeX Data}

Near-infrared spectra of 2MASS 0141$-$4633 were obtained with SpeX 
\citep*{rayner2003} on the 3-m
NASA Infrared Telescope Facility.
The target was first observed on 2004 Sep 05 UT in prism mode 
to provide continuous coverage from 0.7-2.5 $\mu$m
in a single order on a 1024$\times$1024 InSb array. 
Use of the $0{\farcs}5$ slit resulted in a resolving power of ${\lambda}/{\Delta}{\lambda}{\approx}150$. 
For accurate sky subtraction, the target was 
observed in three sets of dithered pairs having 180-second integrations per 
position. On the following night, the target was observed 
in cross-dispersed mode 
to provide higher-resolution spectra over 0.9-2.4 $\mu$m in four orders. 
Use of the $0{\farcs}5$ slit resulted in a resolving power of ${\lambda}/{\Delta}{\lambda}{\approx}1200$.
In this mode the target was observed in four dithered pairs having 300-second
intergrations per position.
Observations of the A0 dwarf HD 8977 were taken on both nights near a similar
airmass and time as the target to provide both telluric correction and flux
calibration. For all observations the
instrument rotator was positioned at the parallactic angle. 
For instrumental calibration internal flat-field and argon lamps were observed immediately after the observation of the A0 star. Standard reductions were employed using the Spextool package version 3.2 \citep{cushing2004,vacca2003} as described in 
\citet{burgasser2004}.

\subsection{Keck NIRSPEC Data}

2MASS 0141$-$4633 was also observed on 2004 Nov 06 UT with the Near-Infrared
Spectrometer (NIRSPEC, \citealt{mclean1998,mclean2000}) on the 10-m 
W.~M.~Keck Observatory. Large fluctuations in relative humidity were noted
during the observations, but sky transparency was otherwise fine.
Using a 1024$\times$1024 InSb array and the spectrograph in low-resolution mode, we selected the grating angle to cover the part 
of the J-band window from 1.14 to 1.36 $\mu$m where the most diagnostic 
near-infrared features lie. Use of the $0{\farcs}38$ slit results in a resolving
power of 
${\lambda}/{\Delta}{\lambda}{\approx}2500$.
The data were obtained in two sets of dithered pairs, with
a 300-second exposure obtained at each position. To measure telluric absorption, the A0 dwarf HD 8977
was observed in a couple of 60-second nodded pairs shortly after the target and close to the same 
airmass; such early-A stars are essentially featureless in this region,
only the star's hydrogen Paschen line at 1.282 $\mu$m needing to be interpolated
over during the telluric correction step. Calibrations consisted of neon and argon arc lamp spectra, a dark frame, and a spectrum of a flat-field lamp.  
Standard reductions using the REDSPEC package were employed, as
described in detail in \cite{mclean2003}.

\section{Analysis}

\subsection{Optical Spectral Features}

\subsubsection{Empirical Comparisons}

The LRIS optical spectrum of 2MASS 0141$-$4633 is displayed in Figure 2 along with a comparison sequence of late-M and early-L dwarf standards from \cite{kirkpatrick1999}. 
Although the gross appearance of the spectrum is clearly that of a late-M
or early-L dwarf, details of the
optical spectral morphology of 2MASS 0141$-$4633 are unlike that of standard M and L dwarfs and in fact look unlike any dwarf seen previously. Specifically, the VO bands from 7334 to 7534 \AA\ and from 7851 to 7973 \AA\ look most like those seen in an L0 dwarf but are considerably stronger than any of the spectra in the standard sequence. Furthermore, the TiO bands at 7053 \AA\ and shortward are very weak or absent, and those at 8206 and 8432 \AA\ have strengths similar to those same bands in an L1 dwarf.

Figure 3 shows details of the same spectra to emphasize how the spectrum
of 2MASS 0141$-$4633 departs from the norm. The alkali lines -- \ion{K}{1}, \ion{Na}{1}, \ion{Rb}{1}, and \ion{Cs}{1} -- are 
noticeably weak compared to the same lines in standard late-M/early-L spectra. Moreover, in addition to being weaker the \ion{K}{1} lines are distinctly narrower; the cores of these lines have full widths at half minima (relative to the local pseudo-continuum) that are 1.2 to 1.7 times smaller than those of the other late-M and early-L dwarfs show in Figure 3. The \ion{Li}{1} doublet is not present either, although
this is not unusual at this resolution and spectral type in the standard
spectral sequence. Hydride bands of FeH and CrH are also weaker than
normal. 

Figure 4 places these anomalies in a different context. Shown for comparison here are the LRIS spectrum of 2MASS 0141$-$4633 compared to
LRIS spectra of the old, L1 dwarf companion GJ 1048B and the late-M giant
IRAS 14303$-$1042 (aka BR 1430$-$1042). In the spectrum of
2MASS 0141$-$4633, the strength of the sodium and
potassium lines falls intermediate between the
strengths seen in the dwarf and the giant. Likewise, the
vanadium oxide strengths fall
intermediate between the high- and low-gravity
objects. The titanium oxide strengths, on the other hand, are very
similar to those of the dwarf and quite unlike those of
the giant. 

The inset in Figure 4 shows one final optical spectral comparison. The 
uppermost
spectrum in the inset is the 2MASS 0141$-$4633 spectrum from 2003 Dec 
24 (UT). Just below that is a poorer signal-to-noise spectrum of 
the same object from a night earlier. (Due to poor sky subtraction at longer
wavelengths, this earlier spectrum has not been considered further in this
paper.) The later spectrum exhibits an H$\alpha$ emission line with a 
pseudo-equivalent width of 10.5$\pm$1.5 \AA. The earlier spectrum shows that
the emission was persistent, the measure in this case being 5.0$\pm$2.5 \AA.
Although these two measurements are only 2$\sigma$ discrepant, it appears that
the pseudo-equivalent width measure in the earlier spectrum is contaminated by a
noise spike at shorter wavelengths, as the width of the line is broader than 
in the later spectrum and offset to shorter wavelengths. Hence we believe 
that these spectra provide strong evidence for variability in the 
H$\alpha$ line itself.

Also shown in this inset is the spectrum (at bottom) of the low-gravity L2 
dwarf G 196-3B, which has a measured \ion{Li}{1} pseudo-equivalent width of
6 \AA\ \citep*{kirkpatrick2001}. However, no discernable \ion{Li}{1} line is 
seen in either spectrum of 2MASS 0141$-$4633 to a level of $\sim$1 \AA.

\subsubsection{Physical Interpretation}

We interpret the spectral features in the optical as follows.
Alkali lines have long been known to be gravity sensitive, in which
case the weakness of these lines in 2MASS 0141$-$4633 can be explained
if this object has a much lower gravity than a typical field L dwarf. The 
narrower appearance of the \ion{K}{1} doublet also points to lower gravity 
because these lines lack the familiar pressure broadened wings seen in 
normal early-L dwarfs. 

At face value the strengths of the oxide bands are harder to interpret, but
nonetheless point to low gravity as well. Both TiO and VO disappear
along the standard late-M/early-L dwarf sequence due to condensation, TiO disappearing first and VO at somewhat colder values of T$_{eff}$ \citep*{kirkpatrick1999}. Lower gravity leads to low atmospheric
pressure which tends to inhibit condensation, in the sense that condensation does not occur until even colder
temperatures are reached\footnote{At their very low
gravities, log(g)$\approx$0, late-M giants never reach temperatures low
enough to trigger the formation of TiO- and VO-bearing
condensates. Because the weakening and
disappearance of these two molecules is the hallmark
of the M/L transition, this explains why there is no
such thing as an ``L giant''.}. (See, for example, Table 2 and Figure 2 of \cite{lodders2002}.) In the case of 2MASS 0141$-$4633 the T$_{eff}$
is low enough that the formation of TiO-bearing condensates has begun
but not low enough to trigger VO-bearing condensation.

As lower gravity can successfully explain all of the spectral peculiarities of 2MASS 0141$-$4633, we propose that this is the physical reason underlying its unique spectral signature. Lower gravity means that this object is lower in mass or larger in radius than a typical field dwarf of the same temperature. Both conditions would be present in a very young brown dwarf because it would still be contracting to its final radius and it would not yet have had the time to cool to the temperatures typical of old field objects of the same mass. 

The lack of lithium in 2MASS 0141$-$4633 may at first consideration seem to contradict our hypothesis of a young age and lower gravity. Many previous studies have used the presence of lithium absorption as an indicator of low mass, as solar metallicity brown dwarfs below a mass of $\sim$60 M$_{Jupiter}$ never burn their initial lithium content. Young brown dwarfs are expected to exhibit lithium absorption whereas older, higher mass objects at the same temperature (spectral type) are not. The L2 dwarf
G 196-3B, shown in the inset of Figure 4, is an example of such a young brown dwarf. In 2MASS 0141$-$4633 alkali lines of \ion{Na}{1}, \ion{K}{1}, \ion{Rb}{1}, and \ion{Cs}{1} (Figure 3) are noticeably weaker than those of normal field late-M and early-L dwarfs. The low gravity hypothesis for our object implies a low mass (see \S4), so it will have retained its nacient lithium content too; however, low gravity also will serve to diminish the \ion{Li}{1} line strength just as it does for the other alkali lines. We propose that the lithium line is not detected at our signal-to-noise and resolution simply because the gravity of 2MASS 0141$-$4633 is uncommonly low. This should serve as a cautionary tale: at very low gravities the strength of the lithium line will decrease markedly, greatly decreasing its usefulness as a test of substellarity.

\subsection{Near-infrared Spectral Features}

\subsubsection{Empirical Comparisons}

The SpeX near-infrared spectrum of 2MASS 0141$-$4633 is shown in Figure 5 along with SpeX spectra of normal field late-M and early-L dwarfs from
\cite{cushing2005}. Most notable in the spectrum of 2MASS 0141$-$4633 is the triangular-shaped appearance of the H-band spectrum, which stands in stark contrast to the more rounded H-band peaks seen in the comparison objects. 
Furthermore, the near-infrared spectrum of 2MASS 0141$-$4633 is much redder than spectra of normal late-M/early-L 
dwarfs. Its color of $J-K_s=1.74{\pm}0.05$ from the 2MASS All Sky Point
Source Catalog is significantly redder than the average color of 
$J-K_s{\approx}1.40$ found for normal, field L0 dwarfs \citep*{kirkpatrick2000}.

Also notable are two deep molecular bands centered near 1.06 and
1.18 $\mu$m that are not present in spectra of the normal
late-M/early-L dwarfs. Figure 6 shows this region in more detail, revealing that both bands are caused primarily by VO. Note also that lines of \ion{Na}{1} and \ion{K}{1} along with bands of FeH are weaker in 2MASS 0141$-$4633 than in the comparison spectra. In this figure the nature of the peculiarities in
the 2MASS 0141$-$4633 spectrum mirrors that seen in its optical spectrum.

A further comparison is shown in Figure 7, which presents the NIRSPEC J-band spectrum of 2MASS 0141$-$4633 along with NIRSPEC spectra of a normal early-L dwarf and a late-M giant. 
As is seen in the optical spectra of Figure 4, the spectrum of 2MASS 
0141$-$4633 looks like a hybrid of a dwarf and a giant. Lines of \ion{K}{1}
are clearly weaker than those in the dwarf, but not nearly as weak as those
in the giant. The broad depression between 1.17 and 1.22 $\mu$m appears to be
caused by two major absorptions: VO, which is strong in the giant and very
weak or absent in the dwarf, and FeH, which is strong in the dwarf and much
weaker or absent in the giant. These two features, VO at shorter wavelengths
and FeH at longer ones, conspire to make a broad V-shaped depression centered
near 1.20 $\mu$m in 2MASS 0141$-$4633. Lastly, the strong absorption bands of
TiO starting near 1.24 $\mu$m in the giant appear to have no analog in either
the normal dwarf or 2MASS 0141$-$4633.

\subsubsection{Physical Interpretation}

As discussed for the optical spectrum, the weakness of the alkali lines and the strength of VO bands in the near-infrared can also be attributed to low gravity. Can a lower gravity also be the cause of the oddly shaped H-band peak and the red J-K$_s$ color? It should be noted that triangular-shaped H-band peaks have been reported previously in the spectra of low-mass brown dwarf candidates in the Orion Nebula Cluster \citep*{lucas2001}. In fact the strangeness of the H-band spectra was used by those authors as proof that their candidates were neither normal foreground nor background stars. Because low-gravity model atmospheres by \cite{allard2000} reproduce the spirit of this feature, this was taken as further evidence of youth.

Those authors did not attempt to explain the physical reason for the change in the H-band pseudo-continuum, but we note that this is probably due to a reduction in H$_2$ collision induced absorption (CIA). 
This feature is created by the collective induced quadrupolar moments on symmetric hydrogen molecules by van der Waals forces in a dense gas. Such perturbations are enhanced in atmospheres of higher pressures and densities. From the equation of hydrostatic equilibrium in stellar atmopsheres, we know that atmospheric pressure scales as $dP/d{\tau} \sim P/{\tau} \sim g/{\kappa}_R$, where $P$ is pressure, $\tau$ is the optical depth, $g$ is gravity, and ${\kappa}_R$ is the Rosseland mean opacity. Photospheres of lower pressure can be produced either through lowering the gravity or increasing the metal content (increasing ${\kappa}_R$). As it is unlikely that 2MASS 0141$-$4633 is a super metal-rich L dwarf -- this would explain only the strength of the VO bands but neither the weakness of the TiO bands or the alkali lines -- the most likely explanation is that H$_2$ CIA is reduced because of lower gravity. 

As \cite{borysow1997} show, H$_2$ CIA produces a very broad feature in the near-infrared. At the temperatures typical of late-M and early-L dwarfs ($\sim$2300 - 2500K) the strongest part of the absorption at JHK bands is located near 2.5 $\mu$m with a gradual weakening toward shorter wavelengths. A diminishing of this absorption would have two main effects: 1) Less absorption at K-band relative to J-band produces a redder J-K$_s$ color, and 2) the H-band peak appears less rounded and more peaky. The latter effect is illustrated -- albeit for a slightly hotter, 2800K object -- in the lower right panel of Figure 5 in \cite{borysow1997}. 

It is noteworthy that another early-L dwarf exhibiting features of low
gravity -- the L2 dwarf G 196-3B (see Figure 4) -- also has an unusually red color. DwarfArchives.org currently
lists 37 objects with well measured types of L2, and the average 2MASS
$J-K_s$ color for these is 1.53. Of these thirty-seven, G 196-3B has the 
reddest color of all, $J-K_s = 2.053{\pm}0.057$. This object is a companion
to a young field M dwarf and has been assigned an age of $\sim$100 Myr 
\citep*{rebolo1998}. We expect the reduced influence of H$_2$ CIA, and
consequently an uncommonly red $J-K_s$ color, to be typical of low-gravity
brown dwarfs such as this one. 

As was the case for the optical spectral features, lower gravity can successfully explain all of the spectral peculiarities of 2MASS 0141$-$4633 in the near-infrared. In the next section we use theoretical spectra to attempt to measure the gravity and temperature of 2MASS 0141$-$4633, which will further allow us to estimate its mass and age. 

\section{Comparison to Model Spectra}

Using data presented in the previous sections, we have pieced together a
high-resolution version of the spectrum covering 0.6 to 2.4 $\mu$m. These
pieces have been assembled using the LRIS optical spectrum and the four orders of
the SpeX cross-dispersed near-infrared spectrum. The four orders all fall
on the same array and have the same relative scaling, so we have merely
normalized this set at its peak flux and then normalized the SpeX prism
spectrum to one at the same wavelength. We then multiplied the LRIS spectrum
by a constant to bring its flux value in agreement with the SpeX prism
spectrum at 0.9 $\mu$m. The scaled LRIS spectrum and SpeX cross-dispersed
spectral orders are plotted together in Fig.\ 8 and overlain on the
full SpeX prism data as a comparison. The internal agreement is excellent.

This combined high-resolution spectrum was then compared to a grid of synthetic
spectra computed with the {\tt PHOENIX} model atmosphere code.  These models
are similar to the ``dusty'' models described in \cite{allard2001}; however,
they include many improvements incorporated in {\tt PHOENIX} since 2001.
Our model grid spanned $T_{eff} = 1500$K to 3000K in 100K increments and
$log(g) = 2.0$ to 6.0 (in cgs units) in 0.5 dex increments.

The effective temperature was obtained by fitting the 0.6-2.4 $\mu$m spectrum
while masking the 1.32-1.60 and 1.75-2.20 $\mu$m regions.  These two regions
correspond to strong H$_2$O absorption bands known to be poorly represented by
current brown dwarf model atmospheres.  A best fit of $T_{eff} = 2000{\pm}100$K 
was determined using a simple least-squares method.  With $T_{eff}$ fixed to 
this value, the gravity was determined by simultaneously fitting the two KI 
doublets in the $J$-band spectrum, again with a least-squares method.  These 
pairs of KI
lines are known to be sensitive gravity indicators (\citealt{gorlova2003},
\citealt{mcgovern2004}) and together yield a best fit of $log(g) = 4.0{\pm}0.5$.
Figure 9 shows these best fits. We note that our 2000K best fit to the entire
SED could also be used to estimate gravity, as the CIA H$_2$ is
a huge contributor to the SED shape at H and K bands. This best fit yields
$log(g) = 3.5$, consistent with the line-fitting method.

Empirically, these derived values roughly agree with expectations. For normal
field early-L dwarfs with measured trigonometric parallaxes, we find $T_{eff}
\approx 2500-2300$K \citep*{kirkpatrick2005araa}. Very late Mira-type M giants
such as our comparison stars IRAS 14303$-$1042 \citep*{wils2003} and IO
Virginis \citep*{kazarovets1993} have types between M8 and M10+
\citep*{kirkpatrick1997}, and the values of $T_{eff}$ for such stars range 
from $\sim$2000 to $\sim$2500K \citep*{vanbelle1996}. Based on this empirical
evidence, we would expect that the effective temperature of 2MASS 0141$-$4633
would also fall intermediate between the giant and dwarf temperatures ranges;
i.e., $\sim$2200K, or somewhat warmer than the model-fit estimate of
$T_{eff}=2000{\pm}100$K.  The empirical guess should only be used as a crude
guide, but it at least shows that the value from model fits is plausible.

As for the value of $log(g)$, we would expect a value lying intermediate
between that of a dwarf ($log(g) \approx 5$) and a giant ($log(g) \approx 0$),
but probably closer to that of a dwarf since overall the spectral
features are a better match to the former than the latter. So the model-fit
estimate of $log(g)=4.0{\pm}0.5$ also agrees with expectations.

Using these model-fit estimates of $T_{eff}$ and $log(g)$, we can plot 2MASS
0141$-$4633 on the theoretical $T_{eff}$ vs.\ $log(g)$ plane to deduce mass 
and age. Figure 10 shows this plane overplotted with the Dusty evolutionary 
tracks from \cite{baraffe2002} and the 1$\sigma$ error box for 2MASS 
0141$-$4633. This suggests that 2MASS 0141$-$4633 has an age between 1 and 50 Myr and a mass between 6 and 25
M$_{Jup}$. The most likely values give an age between 5 and 10 Myr and a mass
of $\sim$12 M$_{Jup}$. Plotting the values of $T_{eff}$ and $log(g)$ on an
alternative set of evolutionary isochrones such as those given by 
\cite{burrows1997} result in identical estimates of $<$50 Myr for the age and 
$<$25 M$_{Jup}$ for the mass, with best guesses of 5 Myr and 11 M$_{Jup}$.

It should be noted, however, that the theoretical models are completely
untested in this regime of mass and age. Any derived values for these 
quantities should be treated with utmost caution. At the moment they are 
useful only in comparing these derived values (in a relative sense) to 
derivations for
young, low-mass brown dwarfs detected elsewhere.  The absolute values of these
quantities are largely untestable at present, but we can nonetheless compare the age
estimate to other empirical data. Based on a grid of late-M and early-L dwarfs
that we are developing for a larger project (Kirkpatrick et al.\, in prep.), 
we can already argue that the age of 2MASS 0141$-$4633 must be substantially
younger than Pleiades members of similar type because its spectral features
belie an even lower gravity. As the Pleiades are believed on other grounds to
be of age $\sim$125 Myr \citep*{stauffer1998}, this means that 2MASS
0141$-$4633 is substantially younger than that value. The fact that the triangular peak in the H-band spectrum is so similar to that of brown dwarf candidates in the Orion Nebula Cluster (age $\approx$ few$\times$Myr) is a further clue that the age is substantially younger than the Pleiades.
Hence, the theoretically
derived best-guess age of 5-10 Myr for 2MASS 0141$-$4633 is realistic.

\section{The Nature of 2MASS 0141$-$4633}

\subsection{Is It a Member of a Young Association?}

Up till now we have considered 2MASS 0141$-$4633 to be a field object,
but is this seemingly solivagant L dwarf actually part of one of the
recently recognized young, nearby, southern associations? (See review by 
\citealt{zuckerman2004}.) An upper age estimate of 50 Myr opens up the 
possibility that
this object is a member of either the AB Doradus Moving Group (age $\approx$
50 Myr\footnote{\cite{luhman2005} argue that the age of AB Dor is closer
to that of the Pleiades, $\sim$125 Myr.}), 
the Tucana-Horologium Association (age $\approx$ 30 Myr), the $\beta$
Pictoris Moving Group (age $\approx$ 12 Myr), or the TW Hydrae Association
(age $\approx$ 8 Myr). 
The best-guess age estimate of 5-10 Myr gives a best match to either the
$\beta$ Pic group or the TW Hya Association. The location of 2MASS 
0141$-$4633 on the sky most closely matches the core of the Tucana/Horologium group or
the $\beta$ Pic group, but
this is not a very strong constraint because these associations are relatively
nearby and hence spread over a large angular extent of sky.

Distance can potentially be used as another crude criterion. 
Using the best-guess estimate of 5-10 Myr for the age and a best guess mass
of $\sim$12 M$_{Jup}$, we find a luminosity $L/L_{\odot} \approx -3.5$ for
2MASS 0141$-$4633 \citep*{burrows1997}. Assuming that the K-band bolometric correction for early-L
dwarfs of 3.2 \citep*{golimowski2004} holds for this object, we find that
the most likely distance, if 2MASS 0141$-$4633 is single, is $\sim$35 pc.
Unfortunately, this also fails to provide much of a constraint.
The median distances of proposed members is found to be 
29 pc for the AB Dor Group,
47 pc for the Tucana/Horologium Association,
39 pc for the $\beta$ Pic group, and
54 pc for the TW Hya Association \citep*{zuckerman2004}.
The proposed members tend to scatter over distances of at least 20 pc
either side of these medians, meaning that the 35-pc distance estimate to
2MASS 0141$-$4633 sheds very little light on possible membership.

Another valuable piece of evidence is the true
space velocity. Astrometric measurements are currently underway
by one of us (CGT), and these will provide a trigonometric parallax and a 
tangential velocity. An accurate radial velocity will be required to determine the
true space velocity and to compare the $(U,V,W)$ space motions to those of the
aforementioned associations. The mean $(U,V,W)$ motions of these groups are,
not surprisingly, very similar (see Table 7 of \citealt{zuckerman2004}), but
a measurement for 2MASS 0141$-$4633 will at least test the hypothesis of
youth in general and the possibility that it is a member of any of these.
Using our highest resolution spectrum (the NIRSPEC J-band spectrum) we derive
a crude radial velocity of 12$\pm$15 km/s for 2MASS 0141$-$4633. This value
is smaller than the total space velocity ($\approx20-30$ km/s) for the 
four young associations mentioned above, so there is no evidence yet to rule
out membership.

Given the scant information currently available, the most likely match to
the age, distance, and sky location of 2MASS 0141$-$4633 is either the 
Tucana/Horologium Association or the $\beta$
Pic Moving Group. Additional observations are clearly warranted.

\subsection{What is its Spectral Type?}

Given that 2MASS 0141$-$4633 is so different from the standard spectral
sequence, its classification should include the fact that it is spectroscopically peculiar. Given that it best matches the spectral features of an early-L dwarf, we will tentatively assign it a spectral type of L0 pec. We predict that 2MASS 0141$-$4633 will begin to look less peculiar as other low-temperature, low-gravity objects are found. Therefore, an augmentation of the L and T spectral classification system will soon be warranted. 

As pointed out in \cite{kirkpatrick2005araa}, we are quickly heading toward a need for other dimensions in our classification of L and T dwarfs to capture information about metallicity and gravity, the two other main parameters besides temperature and clouds that modulate the appearance of the emergent spectrum. We suggest a nomenclature in which the prefix ($d$, $sd$, $esd$) indicates the metallicity type, the core indicates the temperature/cloud type (L0, L1, L2... T0, T1, etc.), and the suffix (perhaps $\alpha$, $\beta$, $\gamma$, etc.) indicates the gravity type. As usual we intend to follow the precepts on the MK Process \citep*{garrison1984} and base our system on only empirical constraints. Normal field objects of solar metallicity would earn a prefix of $d$. Mildly metal deficient L and T dwarfs would earn the prefix $sd$ and extremely metal poor dwarfs would earn the prefix $esd$, following the nomenclature of \cite{gizis1997}. The gravity index is actually a measure of the change in line/feature strength at various ages. Old L or T dwarfs in the field would have a suffix of $\alpha$, those L (and T) dwarfs of Pleiades age would earn a suffix of $\beta$, objects roughly of TW Hya age would get a designation of $\gamma$, and the even younger objects, say of the age of the Orion Nebula Cluster, would get a $\delta$. Following this reasoning, 2MASS 0141$-$4633 might eventually be tagged with a type such as dL0$\gamma$ or dL0$\delta$ to more fully incorporate its spectral appearance into the spectral classification itself.

Establishing this spectral nomenclature scheme is the subject of an on-going program.  Building an empirical spectroscopic grid in metallicity (Burgasser et al.\, in prep.) and gravity (Kirkpatrick et al.\, in prep.) phase space is the topic of future papers. 

\section{Conclusions}

We have presented the discovery of 2MASS J01415823$-$4633574, an odd early-L dwarf whose peculiarities we prescribe to unusually low gravity. Both empirical and theoretical considerations point to a very young age, likely in the 5-10 Myr range. Fits to model atmospheres suggest an effective temperature of 2000$\pm$100 K, and placement on theoretical evolutionary tracks indicate a mass of roughly 12 M$_{Jup}$. Its southerly declination, young age, and implied mass suggest that this may be a very low-mass brown dwarf member of a young association such as the Tucana/Horologium Association or the $\beta$ Pic Moving Group. Other detailed observations of this object -- including spectroscopic acquisition at higher resolution or at longer wavelengths, astrometric and radial velocity measurements, and imaging searches for even fainter companions -- are all important follow-up studies needed to understand better the nature of this rare L dwarf.

\section{Acknowledgments}

We would like to thank our Keck observing assistants Steven Magee, Terry Stickel, and Cynthia Wilburn and our IRTF observing assistant Paul Sears for providing expert operation of the telescopes during our runs. We are also indebted to our instrument scientists, Paola Amico, Grant Hill, and Marc Kassis at Keck and John Rayner at IRTF for their expertise in running the spectrographs, and to Barbara Schaefer and Popoki for making our Christmas runs at Keck always so pleasant. JDK would like additionally to thank the Watkins Foundation at Wichita State Univesity for awarding him a Watkins Visiting Professorship in 2003 during which both the groundwork of this paper was laid and the collaboration with TSB was begun. He also acknowledges very fruitful discussions with Dave Alexander and Jason Ferguson during this trip. We thank Ben Zuckerman, Adam Burrows, and Michael Cushing (the referee) for other suggestions and advice. This paper uses data from the Brown Dwarf Spectroscopic Survey Archive (\url{http://www.astro.ucla.edu/~mclean/BDSSarchive}), the IRTF Spectral Library (\url{http://irtfweb.ifa.hawaii.edu/~spex/spexlibrary/IRTFlibrary.html}), and \url{http://DwarfArchives.org}. This publication also makes use of data products from the Two Micron All Sky Survey, which is a joint project of the University of Massachusetts and the Infrared Processing and Analysis Center/California Institute of Technology, funded by the National Aeronautics and Space Administration and the National Science Foundation. This research has also made use of the NASA/IPAC Infrared Science Archive, which is operated by the Jet Propulsion Laboratory, California Institute of Technology, under contract with the National Aeronautics and Space Administration. As all three flavors of spectra were obtained from the summit of Mauna Kea, the authors wish to recognize and acknowledge the very significant cultural role and reverence that this mountaintop has always had within the indigenous Hawaiian community.  We are most fortunate to have the opportunity to conduct observations there.

\noindent 

\begin{figure}
\epsscale{1.0}
\plotone{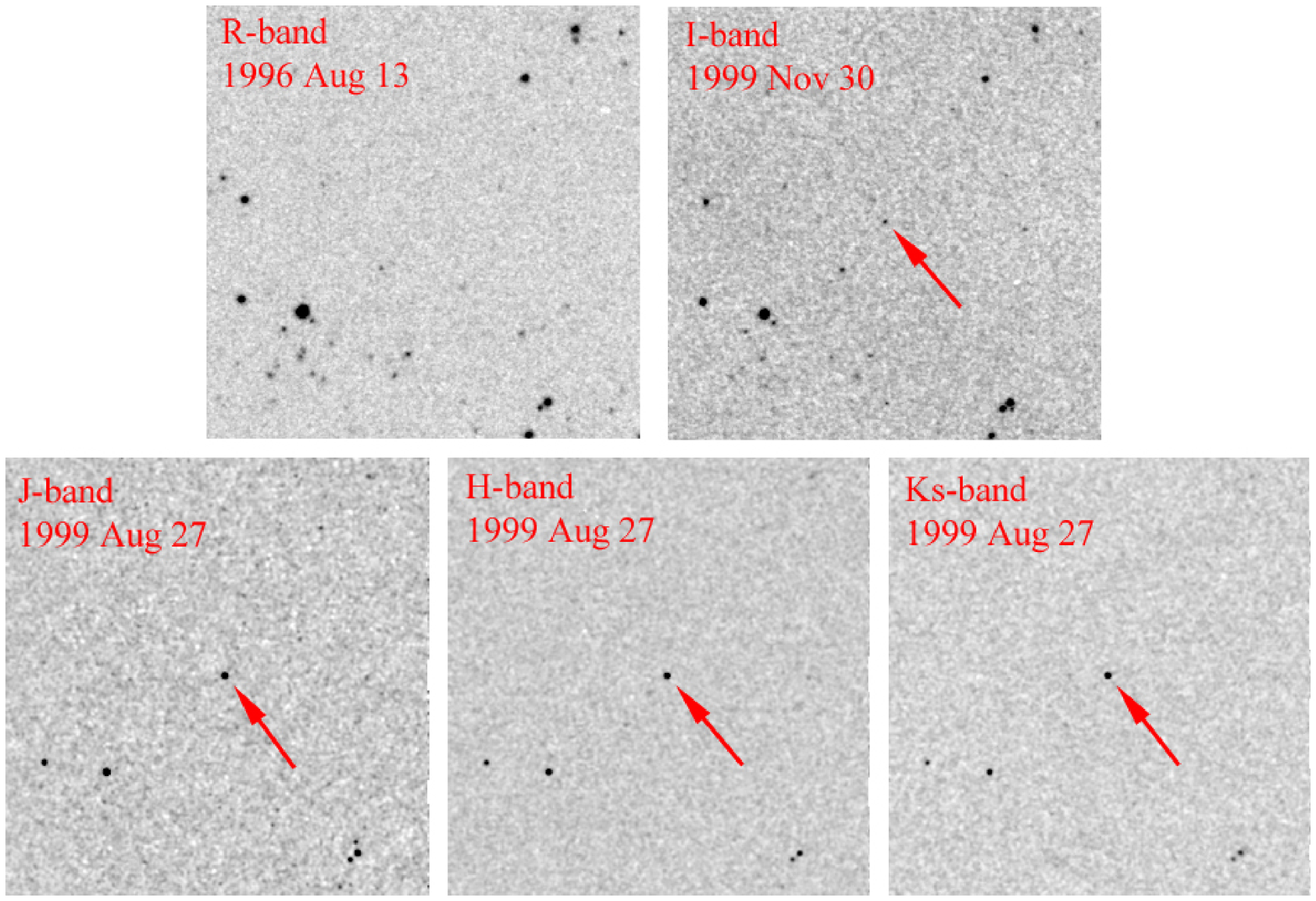}
\caption{Finder charts for 2MASS J01415823$-$4633574. Each field is centered
on the target and is 5 arcminutes wide with North up and East to the left.
The top two images come from the Digitized Sky Survey and the bottom three
images from the Two Micron All Sky Survey. Bandpasses and observing epochs (in UT)
are noted on each image. The object is not seen at R-band, but an arrow marks
its position in the other four. \label{fig1}}
\end{figure}

\clearpage

\begin{figure}
\epsscale{0.65}
\plotone{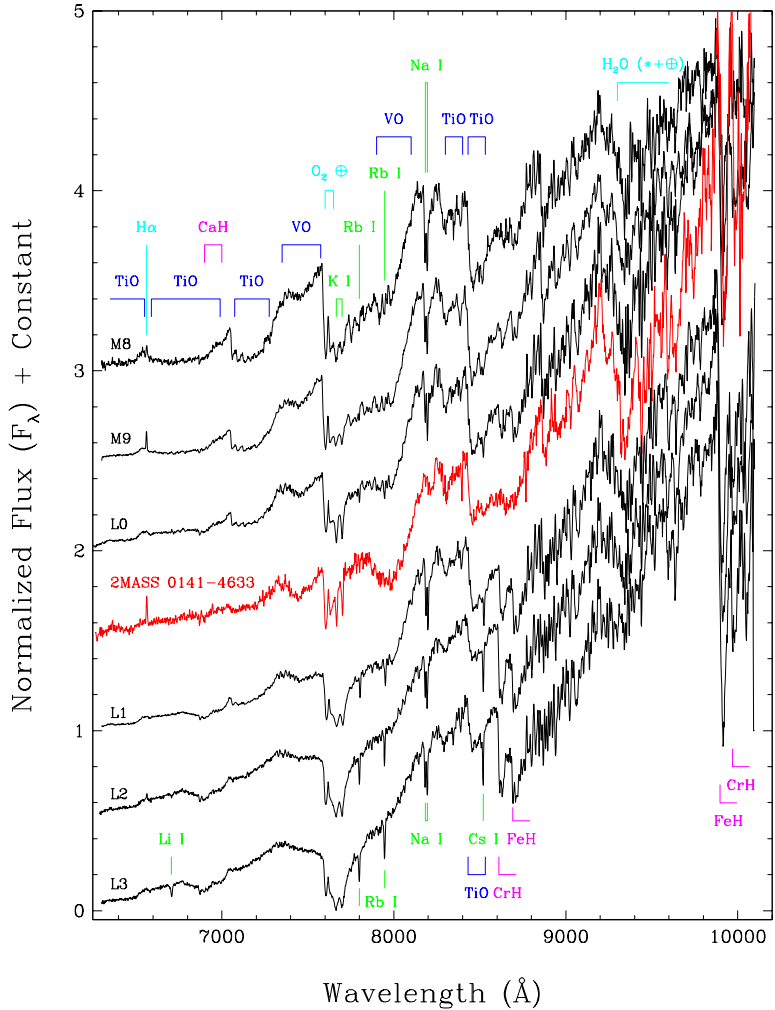}
\caption{The far red optical spectrum of 2MASS
J01415823$-$4633574 (red) compared to a sequence of
normal field late-M and early-L dwarfs (black). The plotted standards
are 2MASSW J1434264+194050 (M8),
2MASSW J1239194+202952 (M9),
2MASP J0345432+254023 (L0),
2MASSW J1439284+192915 (L1),
Kelu-1 AB (L2),
and 2MASS J1146345+223053AB (L3),
as plotted in Figure 6 of \cite{kirkpatrick1999}. All
spectra were taken with the Low-Resolution Imaging
Spectrograph (LRIS) at the 10-m W. M. Keck
Observatory and none of these have been corrected for telluric absorption.
The flux scale is in units of $F_\lambda$ normalized to one at 8250 \AA. The spectra are separated by half-integral offsets along the y-axis for
clarity. Atomic absorption by alkali metals
(green), and molecular absorption by oxides (blue)
and hydrides (magenta) are marked. Telluric
absorption and intrinsic H$\alpha$ emission are also marked
(cyan). Note the weak or absent TiO bands, strong VO bands,
weak alkali lines, weak FeH and CrH bands, and notable H$\alpha$ emission in the
spectrum of 2MASS 0141$-$4633. \label{fig2}}
\end{figure}

\clearpage 

\begin{figure}
\epsscale{0.75}
\plotone{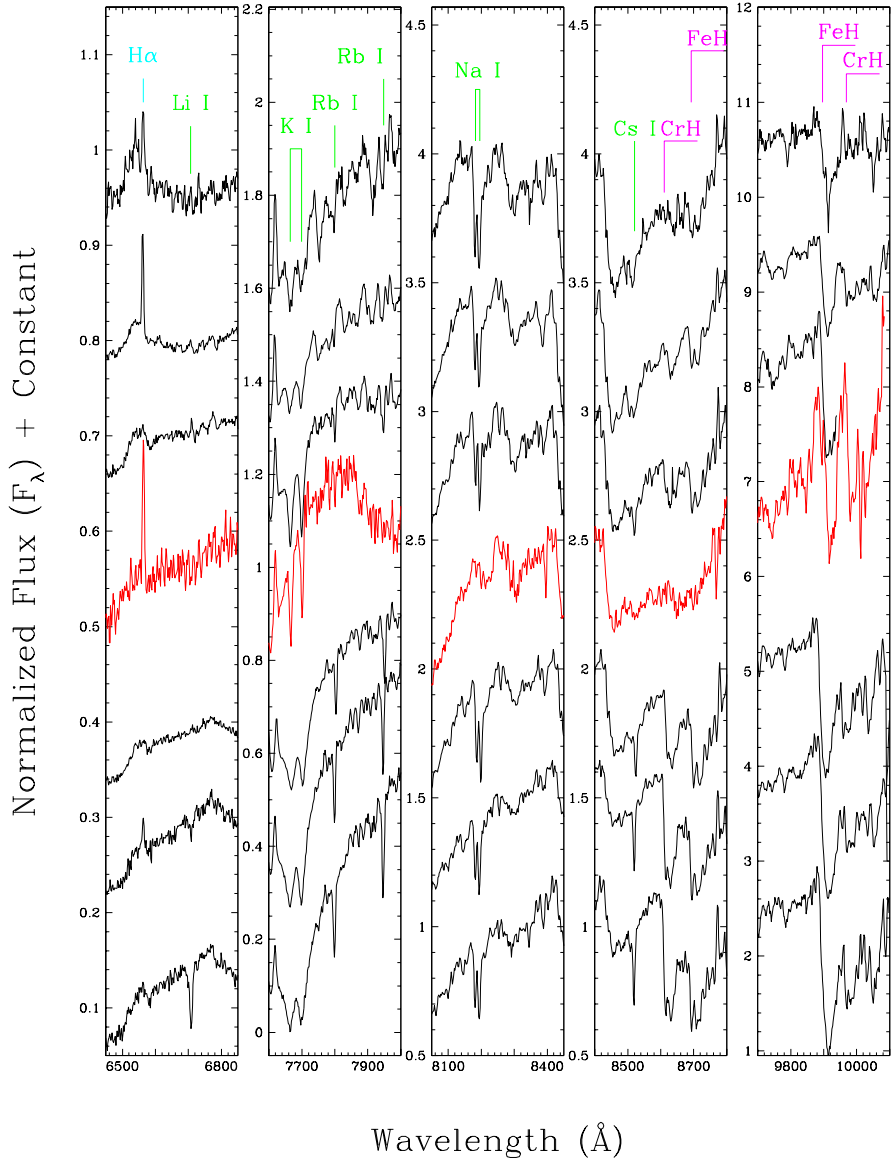}
\caption{Detailed slices of the spectral sequence shown in Fig.~2. Vertical offsets between spectra have been adjusted to ease comparison; vertical adjustments are in quantized units of 0.15 for the leftmost panel, then, moving toward the right, 0.25, 0.5, 0.5, and 1.5 for subsequent panels.
The ordering from top to bottom and the color coding are the same as in 
Figure 2. As noted before, \ion{K}{1}, \ion{Na}{1}, \ion{Rb}{1}, and 
\ion{Cs}{1} lines in
2MASS 0141$-$4633 are 
noticeably weaker than those of standard late-M and early-L dwarfs. (The other
alkali line, \ion{Li}{1} is not seen in 2MASS 0141$-$4633.) Moreover, the cores of the \ion{K}{1} 
doublet are also much narrower in 2MASS 0141$-$4633. Hydride bands of
FeH and CrH are also weaker and H$\alpha$ emission prominent in
the spectrum of 2MASS 0141$-$4633. \label{fig3}}
\end{figure}

\clearpage 

\begin{figure}
\epsscale{0.60}
\plotone{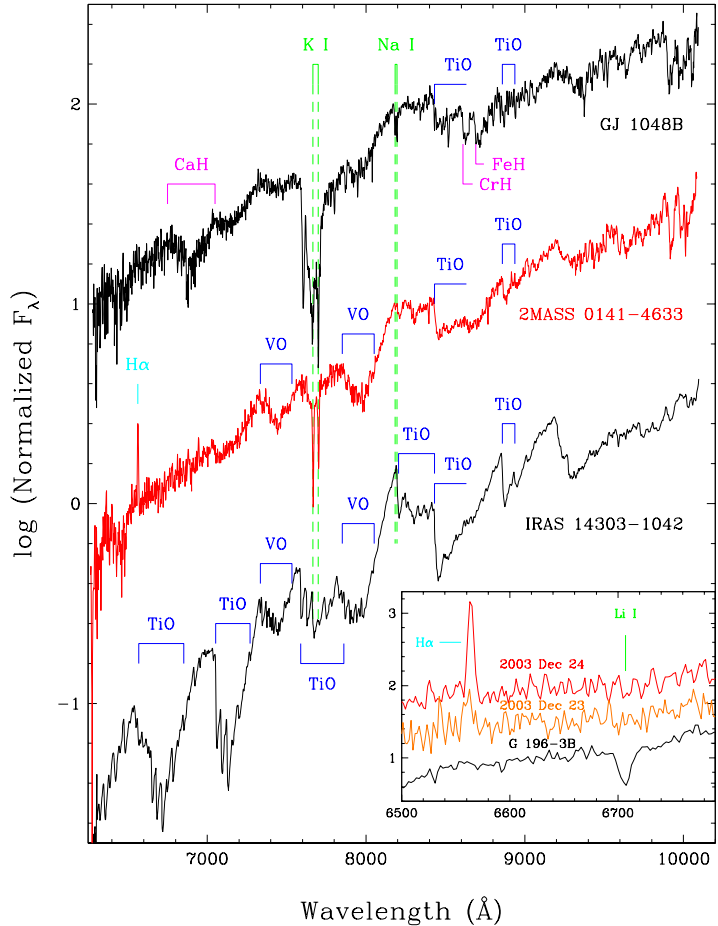}
\caption{Far red optical spectra of a normal L1 dwarf GJ 1048B (top, black), the peculiar L dwarf
2MASS 0141$-$4633 (center, red), and the very late type
M giant IRAS 14303$-$1042 (bottom, black). All
spectra were taken with LRIS on Keck and in this case all spectra
have been telluric corrected. Spectra have been normalized to one at 8250 \AA\ and separated vertically by 1 dex for ease of comparison. Color coding
of features is the same as in Fig.~2. In the spectrum of 2MASS 0141$-$4633, the strength of the sodium and potassium lines, as well as the strength of the vanadium oxide bands, falls intermediate between the
strengths seen in the dwarf and the giant. The titanium oxide strengths, on
the other hand, are very similar to those of the dwarf and quite unlike those of the giant. The inset shows an expanded version of the 2MASS 0141$-$4633
spectrum (top, red) along with a spectrum of the same object taken one night earlier (middle, orange). Note the change in H$\alpha$ emission strength between the two nights. Also shown in the inset for comparison is the low gravity L2 dwarf G 196-3B. Unlike G 196-3B, the spectrum of 2MASS 0141$-$4633 shows no indication of \ion{Li}{1} absorption. All three spectra in the inset have been normalized to unity at 6650 \AA, displayed in linear flux units, and separated vertically from one another by units of 0.5.
\label{fig4}}
\end{figure}

\clearpage 

\begin{figure}
\epsscale{0.7}
\plotone{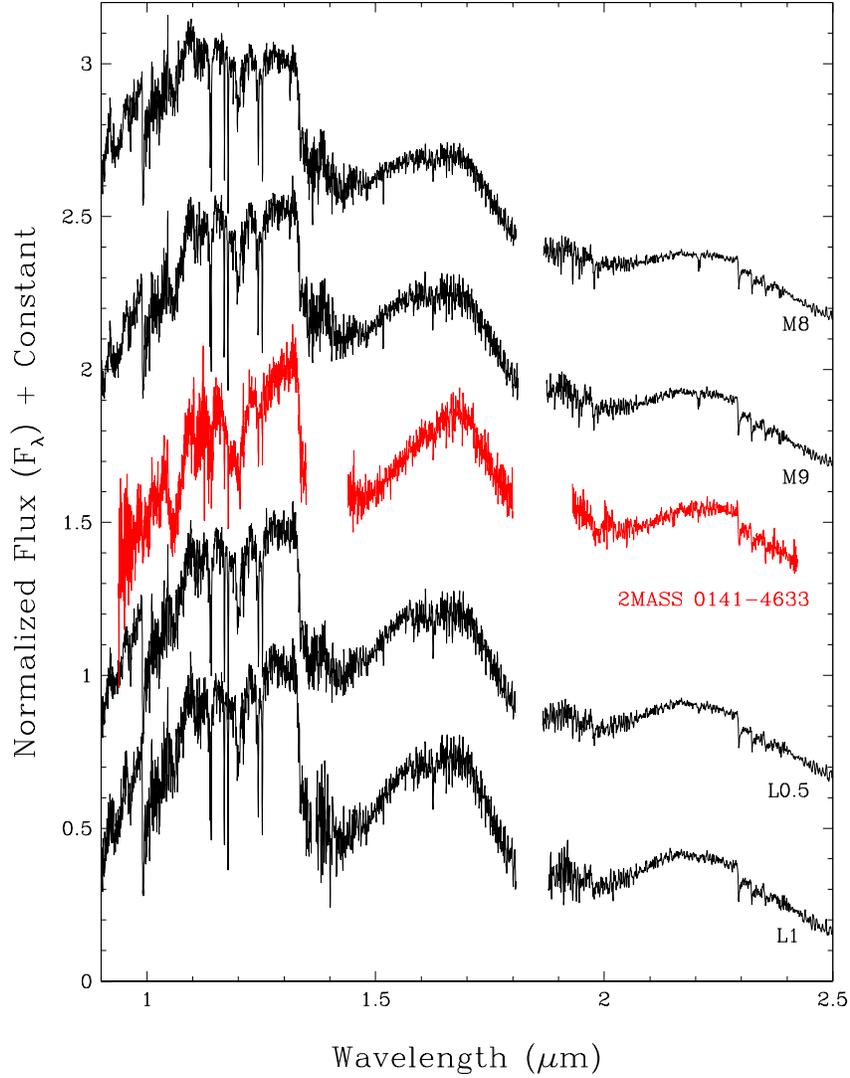}
\caption{Near-infrared spectra of normal field late-M
and early-L dwarfs (black) and the peculiar L
dwarf 2MASS 0141$-$4633 (red). The comparison spectra are
vB 10 (M8), LHS 2924 (M9), 2MASSI J0746425+200032AB (L0.5), and 2MASSW J1439284+192915 (L1) from \cite{cushing2005}. All spectra were
taken with SpeX on the 3-m NASA Infrared
Telescope Facility and have been normalized to one at 1.30 $\mu$m. 
Spectra have been separated vertically in units of 0.5 for clarity. Most notable in the spectrum of
2MASS 0141$-$4633 is the triangular-shaped
appearance of the H-band spectrum. Also notable
are two deep bands centered near 1.06 and
1.18 $\mu$m that are not present in spectra of the normal
late-M/early-L dwarfs. As can also been seen in this
figure, the near-infrared spectrum of 2MASS 0141$-$4633 
is much redder than spectra of normal late-M/early-L 
dwarfs. \label{fig5}}
\end{figure}

\clearpage 

\begin{figure}
\epsscale{0.75}
\plotone{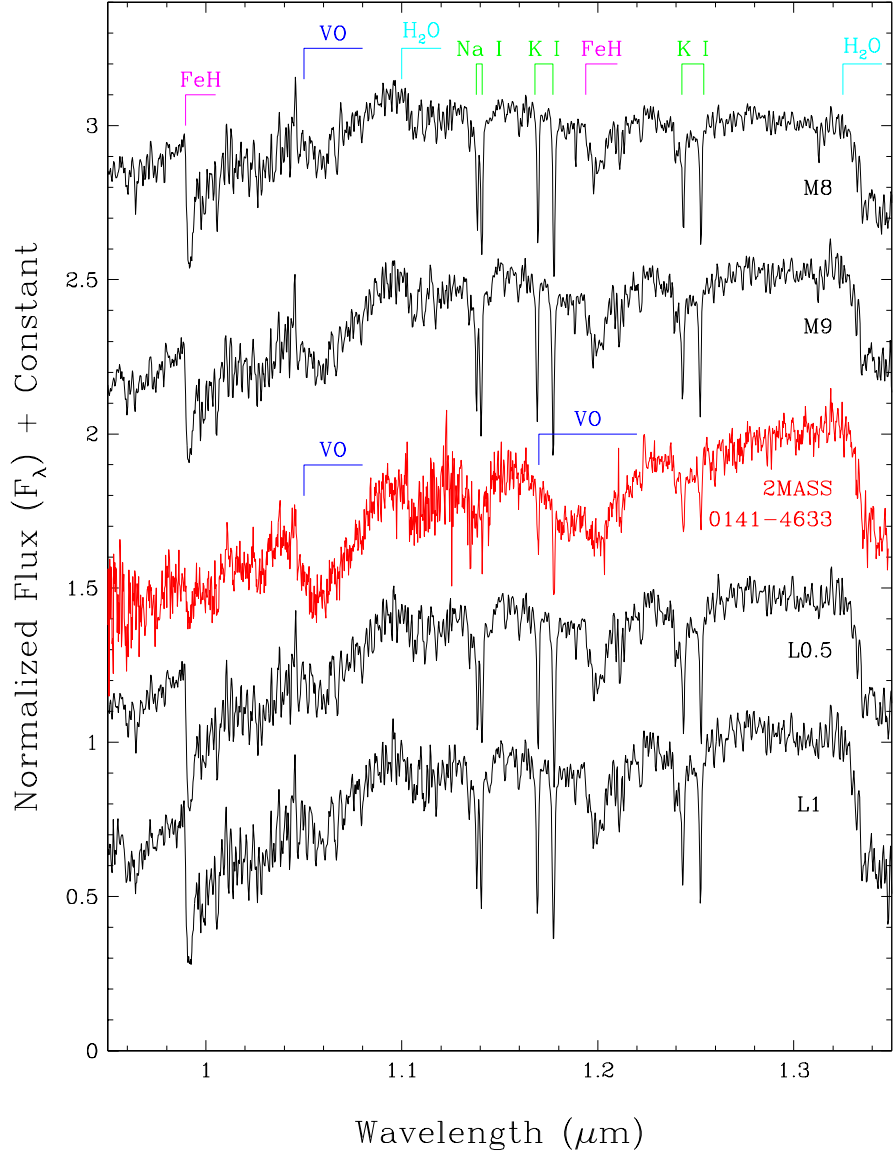}
\caption{A detail of the J-band spectra from Fig.~5. The ordering and
color coding are the same as in that figure. Compared to the sequence of
normal late-M and early-L dwarfs, the spectrum of 2MASS 0141$-$4633 shows alkali lines (green) and FeH bands
(magenta) that are weaker than normal and VO bands (blue) that are stronger than normal, 
mirroring what is seen in the optical spectra. \label{fig6}}
\end{figure}

\clearpage 

\begin{figure}
\epsscale{0.65}
\plotone{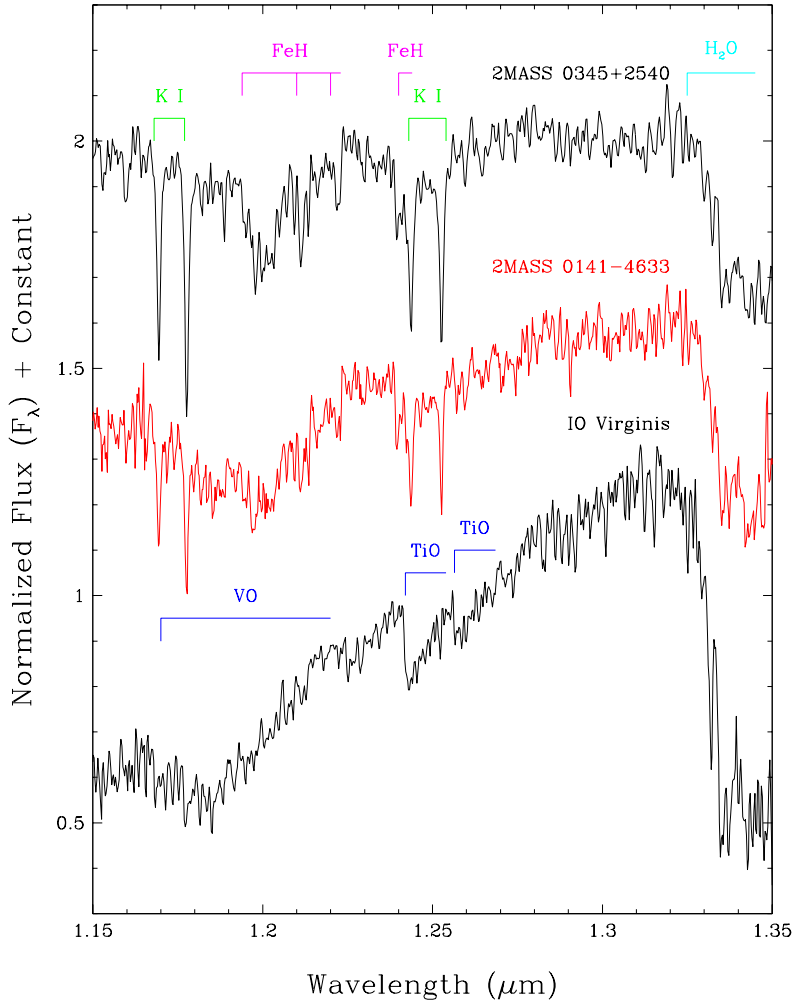}
\caption{The Keck-NIRSPEC spectrum of 2MASS 0141$-$4633 compared to NIRSPEC spectra of a normal
field L0 dwarf (2MASP J0345432+254023, from \citealt{mclean2003}) and a late-type M giant (IO Virginis, from \citealt{mcgovern2004}).
Spectra have been normalized to one at 1.27 $\mu$m and offset by half integers
to separate the spectra vertically. Color coding of features is the same as
in Fig.\ 2. As with Fig.\ 4 the spectrum of 2MASS 0141$-$4633 in this 
wavelength regime shows features
intermediate between the dwarf and the giant. Potassium lines are considerably
weaker than those seen in the dwarf, but not as weak as the practically
non-existent lines in the giant. The broad trough between 1.17 and 1.22 $\mu$m
is caused by a combination of features: a deep vanadium oxide band (strong in 
the giant) on the short wavelength side and a deep series of iron hydride
bands (strong in the dwarf) on the long wavelength side. Bands of titanium
oxide situated longward of 1.24 $\mu$m, which are seen in the giant, 
appear not to be present in either the normal L0 dwarf or the spectrum of 
2MASS 0141$-$4633.
\label{fig7}}
\end{figure}

\clearpage 

\begin{figure}
\epsscale{0.7}
\plotone{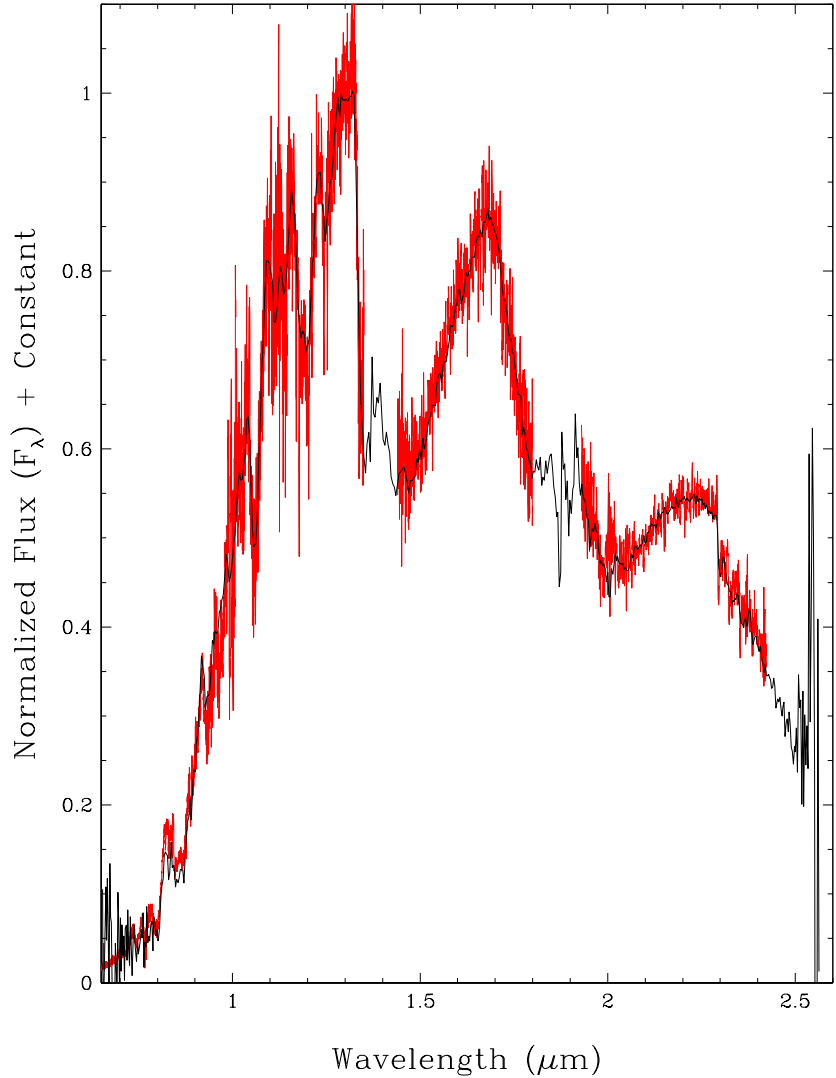}
\caption{A plot of the full spectral energy distribution of 2MASS 0141$-$4633
from 0.6 to 2.5 $\mu$m. The data in black represent the Keck-LRIS optical 
spectrum
from Fig.~2 along with the IRTF-SpeX cross-dispersed spectrum from Fig.~5.
Overplotted in red is the IRTF-SpeX prism spectrum. The SpeX prism spectrum was normalized to one at its peak flux (at 1.32 $\mu$m). The SpeX cross-dispersed spectrum was normalized to one at this same wavelength and the LRIS spectrum multiplied by a constant to match the prism spectrum at 0.9 $\mu$m. There is excellent agreement in all overlaps -- between the SpeX prism and cross-dispersed data, the SpeX prism and LRIS data, and even in the 0.94-1.01 $\mu$m overlap between the short-wavelength order of the SpeX cross-dispersed spectrum and the LRIS spectrum.
\label{fig8}}
\end{figure}

\clearpage 

\begin{figure}
\epsscale{0.7}
\plotone{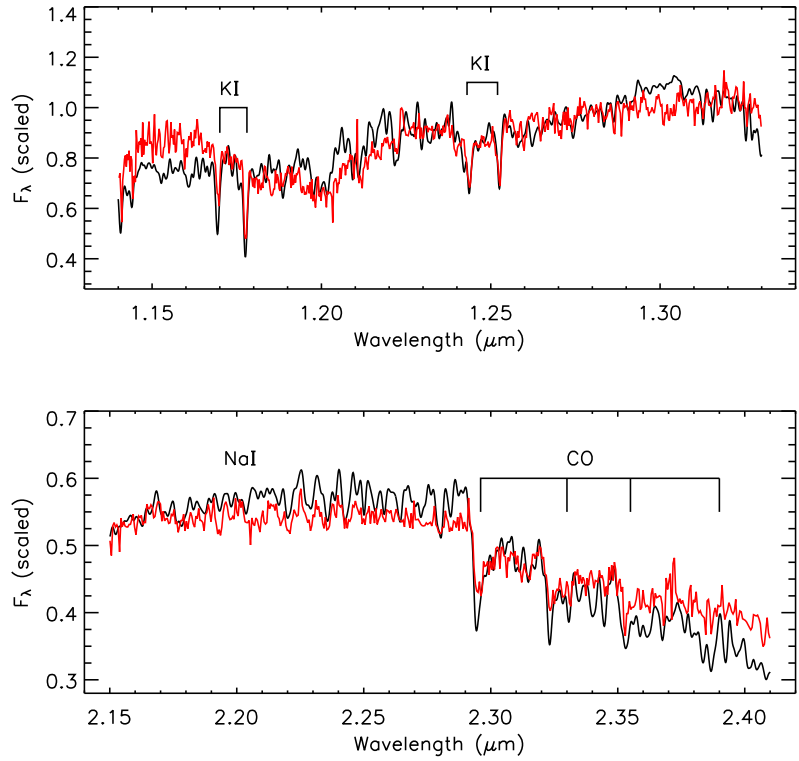}
\caption{$J$-band (top) and $K$-band (bottom) spectra for 2MASS 0141$-$4633.  
The observed spectrum is shown in red, and the
best fitting brown dwarf model spectrum (with $T_{eff} = 2000$K and 
$\log(g) = 4.0$)
is plotted in black.  The gravity sensitive \ion{K}{1} lines are indicated in the
$J$-band as well as the weak \ion{Na}{1} doublet and the CO features in the
$K$-band. 
\label{fig9}}
\end{figure}

\clearpage 

\begin{figure}
\epsscale{0.7}
\plotone{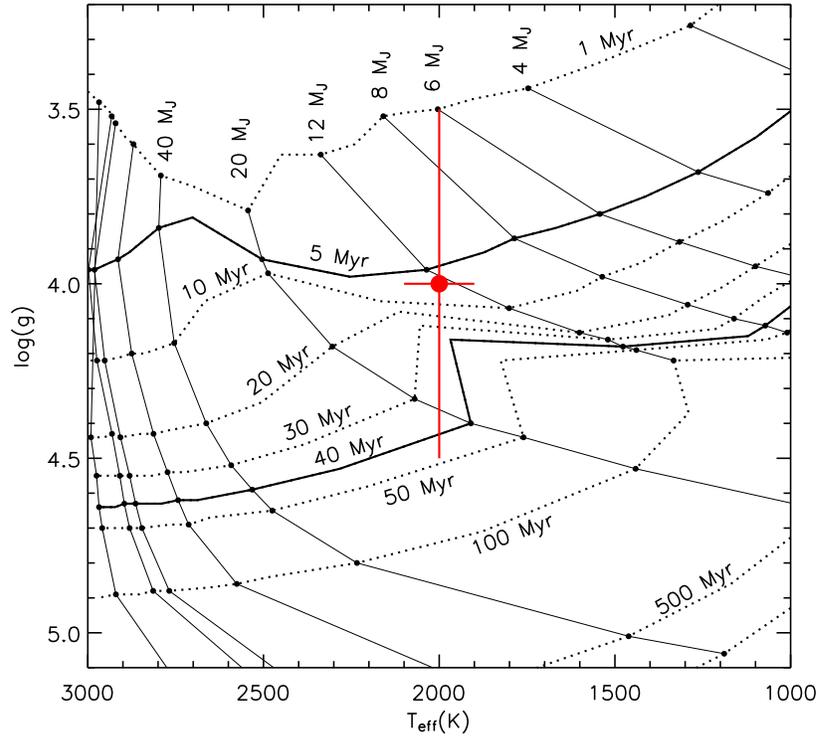}
\caption{Theoretical $T_{eff}$ --\ $\log(g)$ evolutionary tracks from \citep*{baraffe2001}
and \citep*{chabrier2000}. Isochrones are plotted as dotted lines while thin
solid lines indicate constant mass.  Isochrones for 5 and 40 Myr have been
highlighted with thick solid lines.  The filled circle with error bars marks
the best fit $T_{eff}$ and $\log(g)$ values for 2MASS 0141$-$4633 deduced from
brown dwarf model spectra.  With these values, the evolution models imply
an age between 1 and 50 Myr and a mass between 6 and 25 M$_{Jup}$.
\label{fig10}}
\end{figure}

\end{document}